Ferromagnetic van der Waals compound MnSb$_{1.8}$Bi$_{0.2}$Te$_4$

Yangyang Chen[1,3], Ya-Wen Chuang[1], Seng Huat Lee[1,2], Yanglin Zhu[1], Kevin Honz[1], Yingdong Guan[1], Yu Wang[1], Ke Wang[5], Zhiqiang Mao[1,2] and Jun Zhu[1]

Colin Heikes[4], P. Quarterman[4]
Pawel Zajdel[6], Julie A. Borchers[4]
William Ratcliff II[4,7]

[1]Department of Physics, The Pennsylvania State University, University Park, Pennsylvania 16802, USA

[2]2DCC, The Pennsylvania State University, University Park, Pennsylvania 16802, USA

[3] International Center for Quantum Materials, School of Physics, Peking University, Beijing 100871, PR China

[4]NIST Center for Neutron Research, NIST, Gaithersburg, MD 20899, USA

[5]Materials Characterization Laboratory, Materials Research Institute, The Pennsylvania State University, University Park, Pennsylvania 16802, USA

[6] Institute of Physics, University of Silesia, ul. 75 Pulku Piechoty 1, 41-500, Chorzow, Poland

[7]Department of Materials Science and Engineering, University of Maryland, College Park, MD 20742, USA

Correspondence to: jxz26@psu.edu

**Abstract**

The intersection of topology and magnetism represents a new playground to discover novel quantum phenomena and device concepts. In this work, we show that a van der Waals compound MnSb$_{1.8}$Bi$_{0.2}$Te$_4$ exhibits a ferromagnetic ground state with a Curie temperature of 26 K, in contrast to the antiferromagnetic order previously found for other members of the Mn(Sb, Bi)$_2$Te$_4$ family. We employ magneto-transport, bulk magnetization and neutron scattering studies to illustrate the magnetic and electrical properties of MnSb$_{1.8}$Bi$_{0.2}$Te$_4$ and report on the observation of an unusual anomalous Hall effect. Our results are an important step in the synthesis and understanding of ferromagnetic topological insulators.

A central theme of contemporary condensed matter research explores the notion of topology and symmetry to generate novel quantum phenomena [1]. A good example is the quantum anomalous Hall effect (QAHE) found in magnetic topological insulators (TI) Cr$_{0.15}$(Bi, Sb)$_{1.85}$Te$_3$, where magnetism introduced by Cr doping breaks the time reversal symmetry and gives rise to robust chiral edge states that can carry current ballistically without the need of an external magnetic field [2,3]. Chiral magnetic textures such as skyrmions are another good example [4-6]. The generation of spin chirality requires the breaking of crystalline inversion symmetry, which can be achieved in bulk materials [7-9] or in heterostructures that combine ferromagnetism (FM) with strong spin orbit coupling [10-14]. Magnetism and heterostructures in the van der Waals (vdW) geometry provides another powerful natural platform to explore magnetic anisotropy and interface effect, with the added benefit of electric field control for potential device applications [15-20].

Recent research activities have identified a promising vdW magnetic compound family, Mn(Bi, Sb)$_2$Te$_4$ with the potential for realizing the QAHE above dilution temperatures [21-32]. Mn(Bi, Sb)$_2$Te$_4$ can be regarded as consisting of a Mn-Te layer inserted into the quintuple layer of (Bi, Sb)$_2$Te$_3$ (Fig. 1(a)). Here ordered Mn layers creates magnetism ideally without introducing disorder. Ferromagnetic coupling between adjacent Mn layers would lead to the QAHE, however in the most studied compound of the family, MnBi$_2$Te$_4$, A-type antiferromagnetism (AFM) dominates the interlayer Mn coupling [21-23,30,31,33-35]. Theoretically only odd-layered flakes are expected to exhibit the QAHE [24,25,27-29]. Measurements on such samples, however, show that conductance quantization only occurs at external field of several Tesla [24]. The realization of the QAHE at zero magnetic field remains an ongoing pursuit.

We show here that an Sb-rich composition of the family, MnSb$_{1.8}$Bi$_{0.2}$Te$_4$, can be stabilized into a FM phase with a Curie temperature of $T_C$ = 26 K. We present transport, magnetometry and neutron diffraction measurements that illustrate the properties of this FM state. Our sample also exhibits an unconventional anomalous Hall effect. These findings open up new possibilities of engineering magnetic and topological structures and devices in this vdW ferromagnet.

Crystals of MnSb$_{1.8}$Bi$_{0.2}$Te$_4$ and MnSb$_2$Te$_4$ used for comparison are synthesized by a flux method. X-ray diffraction and scanning transmission electron microscopy (STEM) measurements confirms the rhombohedral structural phase in both materials, with greater than 95% purity. The growth method and characterization are given in Section S1 of the Supplementary Material (SM). Figure 1(a) illustrates a schematic sideview of the layer stacking in the Mn(Bi, Sb)$_2$Te$_4$ family. Micrometer-sized flakes are exfoliated from selected crystals and transferred using a polypropylene carbonate stamp to pre-patterned electrodes inside a glovebox filled with argon gas. The finished device is covered with a droplet of Poly(Methyl MethAcrylate) before transferred to a pumped $^4$He cryostat. Figure 1(b) shows an optical image of a typical Hall bar device. Flakes of similar color tone measure 100-300 nm in thickness in an atomic force microscope. Transport measurements are performed in a pumped $^4$He cryostat with a magnetic field up to 9 T using standard low-frequency techniques. Magnetometry measurements are performed in a SQUID magnetometer from 2-300 K. Single crystal elastic neutron scattering measurements are performed using the BT-4 and BT-7 triple axis spectrometers (TAS) at the NIST Center for Neutron Research (NCNR). Instrumentation details of the neutron studies can be found in Section S5 of Ref. [30].

Figure 1(c) plots the Hall resistance $R_{xy}(H)$ obtained on a MnSb$_{1.8}$Bi$_{0.2}$Te$_4$ device. A small $R_{xx}$ component is removed from the data though an asymmetrization step that averages the upsweep of $R_{xy}(H)$ and the downsweep of $-R_{xy}(-H)$. Measurements were taken at a series of fixed temperatures ranging from 2 to 60 K. Traces plotted here represent the typical behavior in different temperature ranges. As the $T$ = 60 K trace shows, $R_{xy}(H)$ is a straight line from −9 T to 9 T at high temperatures. An anomalous Hall effect starts to develop at $T$ < 46 K, where the slope $dR_{xy}/dH$ taken at $R_{xy}$ = 0 (illustrated by a green dashed line for the $T = 2$ K upsweep trace) becomes larger than the slope taken at high field (a gray dashed line in the inset). The difference of the two originates from a non-zero magnetization ($M$) of the sample since $R_{xy}(H) = R_0H + R_sM$ [36]. At sufficiently high field where $M$ saturates, the slope $dR_{xy}/dH$ yields the normal Hall coefficient $R_0 = 1/ne$. $R_0$ follows a cos $\theta$ dependence as the external field tilts away from the $c$-axis of the crystal (Fig. S3), which confirms the two-dimensional nature of the mobile carriers and yields a hole density of $n_h = 6.3\times10^{15}$/cm$^2$. This translates to a hole doping of ~ $10^{13}$/cm$^2$ per septuple layer and puts the Fermi level in the bulk valence band of MnSb$_{1.8}$Bi$_{0.2}$Te$_4$ [31]. $R_0$ is approximately $T$-independent, as demonstrated in Fig. S3 of the SM. In contrast, the slope $dR_{xy}/dH$ taken at $R_{xy} = 0$, called the zero-field slope from now on, increases rapidly with decreasing temperature and reaches a broad maximum around 12−20 K. Its $T$-dependence is plotted and discussed in Fig. 2(a) below.

In addition to the anomalous Hall effect, $R_{xy}(H)$ becomes hysteretic at temperatures below ~ 25 K. Data at 2 K and 15 K are plotted to show the two different shapes of the hysteresis loop. Hysteresis is also observed in bulk magnetization measurements of the parent MnSb$_{1.8}$Bi$_{0.2}$Te$_4$ crystal. Figure 1(d) plots the $M(H)$ data at $T = 2$ K, from which we extracted a remnant magnetization of $M_0 = 0.6$ μ$_B$/Mn, a coercive field of $H_c = 310$ Oe and a saturated $M$ of 1.8 μ$_B$/Mn (inset of Fig. 1(d)). Both the coercive field $H_c$ and the remnant $R_{xy}$ at $H = 0$ decrease with increasing temperature and vanish at $T > 23$ K (see Fig. S4 of the supplementary material).

The above observations are strong evidence of a ferromagnet, in stark contrast to the AFM ground state found in previous studies of the Mn(Bi, Sb)$_2$Te$_4$ family [21,23,30,31]. In fact, some of our Mn(Bi, Sb)$_2$Te$_4$ samples exhibit the above FM characteristics while others show behaviors similar to the AFM phase reported in Ref. [31] for Sb-rich compositions, despite that all crystals are synthesized via nominally the same growth recipe. The only possible variation in growth condition between the different batches is the quenching temperature and the cooling rate during the quenching process (See Section S1 of the SM for more details). This implies that the FM phase is metastable. Indeed, first principles calculations show that the FM phase competes closely with the AFM phase in the end compound MnSb$_2$Te$_4$ [31,33]. Identifying the precise synthesis conditions that lead to FM order in a wide range of alloy compositions will be an important goal of future studies. In this work, we focus on the properties of MnSb$_{1.8}$Bi$_{0.2}$Te$_4$ samples that display the FM characteristics.

To further explore their magnetic properties, we plot in Figs. 2(a) and 2(b) the $T$-dependent magnetic susceptibility $\chi(T)$, extracted from the zero-field slope of the Hall resistance $dR_{xy}/dH$ and DC magnetometry measurements respectively. In a magnetic system, the zero-field slope $dR_{xy}/dH$ includes the contribution from the out-of-plane magnetic susceptibility $\chi = dM/dH$. In Fig. 2(a), $dR_{xy}/dH$ ascends rapidly at $T \sim 25$ K, reaches a maximum value of 11 Ω/T around 12 - 20 K, which is more than 100 times larger than the normal Hall coefficient $R_0 = 0.1$ Ω/T of this device, before dropping again at lower temperatures. In another word, the zero-field $dR_{xy}/dH$ is dominated by the magnetic response of the system and effectively measures the $\chi(T)$ of the microscope device. The magnetometry studies conducted on bulk crystals tell a similar story. Figure 2(b) plots the temperature-dependent linear susceptibility $M/H(T)$ obtained under both zero-field-cooling (ZFC) and several field-cooling (FC) conditions using several different fields as labeled in the plot. The 50 Oe ZFC data (solid black line) strongly resemble the zero-field $dR_{xy}/dH$ shown in Fig. 2(a), suggesting that our samples behave homogeneously from the µm to the mm length scale. Both support the onset of a FM order at a Curie temperature of $T_c \sim 26$ K. At $T < 12$ K, both the $dR_{xy}/dH$ and the low-field ZFC $M/H$ data show a pronounced drop that deviates from a conventional FM. More complex magnetic phases may emerge in this temperature range [34,37-39]. We aim to understand its nature with additional measurements and analyses [40].

Neutron scattering measurements provide strong evidence supporting an FM order in $MnSb_{1.8}Bi_{0.2}Te_4$. Figures 2(c) and (d) compare results obtained on our $MnSb_{1.8}Bi_{0.2}Te_4$ and $MnSb_2Te_4$ crystals. Upon cooling, the (1 0 1) and (1 0 4) nuclear reflection peaks in $MnSb_{1.8}Bi_{0.2}Te_4$ gained intensity with no peak appearing at the (1 0 2.5) position (Fig. 2(c) inset). In contrast, the (1 0 2.5) peak appeared at low temperatures in our $MnSb_2Te_4$ sample while the amplitude of the nuclear reflections remained unchanged (Fig. 2(d) inset). The (1 0 2.5) peak is associated with the development of the A-type AFM phase in $MnBi_2Te_4$ in previous reports [30,32]. The neutron data clearly indicate a different magnetic order in our samples, that is, AFM in $MnSb_2Te_4$ and FM in $MnSb_{1.8}Bi_{0.2}Te_4$. Mean-field fits to the temperature-dependent scattering amplitude at the (1 0 1) and (1 0 2.5) positions yield a Curie temperature of $T_C \sim 26$ K and a Néel temperature of $T_N \sim 20$ K for the $MnSb_{1.8}Bi_{0.2}Te_4$ and $MnSb_2Te_4$ samples respectively. Further, we show in Fig. 2(a) the zero-field $dR_{xy}/dH$ data we obtained on a $MnSb_2Te_4$ device (solid blue squares). It is consistent with an AFM phase with $T_N \sim 19.5$ K, and is in excellent agreement with previous susceptibility measurements of this material [31].

The sensitivity of $R_{xy}$ to the magnitude of $M$ enables us to determine the saturation field $H_s$ in $MnSb_{1.8}Bi_{0.2}Te_4$ and construct a $H_s$ - $T$ phase diagram. To do this we first determine the anomalous Hall signal $\Delta R_{xy}(H) = R_{xy}(H) - R_0 H$ for the device shown in Fig. 1(b). The results at several temperatures are shown in Fig. 1(e). The saturation field $H_s$ is defined as the field at which the extension of the zero-field slope reaches the saturated value of $\Delta R_{xy}$, as illustrated by the green dashed lines for the $T = 30$ K trace. Figure 3(a) plots the resulting $H_s$ - $T$ diagram. $H_s$ reaches a minimum of 0.016 T near $T_C$. Remarkably, at low temperature, $H_s$ is only 0.043 T in $MnSb_{1.8}Bi_{0.2}Te_4$, in comparison to 0.42 T in $MnSb_2Te_4$ (Fig. S5) and more than 7 T in $MnBi_2Te_4$ (Fig. S6), despite similar ordering temperatures of $\sim 20$ K in all three materials. This observation strongly attests to the FM order in $MnSb_{1.8}Bi_{0.2}Te_4$. The small $H_s$ here is associated with the alignment of the FM domains in an external field, rather than the spin-flop transition of individual Mn moment. In addition, we see that the anomalous Hall effect extends into the paramagnetic phase (open circles in Fig. 3(a)), indicating FM fluctuations are already important at $T \geq T_C$.

Next, we demonstrate the impact of magnetic order on the transport characteristics of $MnSb_{1.8}Bi_{0.2}Te_4$. Figure 3(b) plots $R_{xx}(T)$ traces taken at a series of fixed magnetic fields. We track the sign change of $dR/dT$ as a function of $T$ and $H$ and plot the results on a $H$ - $T$ map, similar to the $H_s$ - $T$ diagram shown in Fig. 3(a). At temperatures above 50 K, $R_{xx}(T)$ exhibits the expected metallic $T$-dependence, i.e. $dR/dT > 0$. As $T$ approaches $T_C$, strong spin fluctuations lead to a slightly insulating $T$-dependence, i.e. $dR/dT < 0$, similar to the situation in $MnBi_2Te_4$ [30]. A positive $dR/dT$ is found again when the moments align spontaneously or under a sufficiently large external field, likely due to the reduction of magnetic scatterings that involve a spin-flip/flop. The onset of another insulator-like regime at $T < 12$ K coincides with the drop of $\chi$ in Fig. 2(a), the nature of which will be elucidated in future studies.

Figure 3(c) plots the normalized magnetoresistance (MR) of a $MnSb_{1.8}Bi_{0.2}Te_4$ device at selected temperatures. In $MnSb_{1.8}Bi_{0.2}Te_4$, MR is always negative. Its magnitude increases with decreasing temperature and becomes hysteretic with the onset of the FM order. Alignment of all moments in an external field results in a large reduction of nearly 20% at low temperature. Comparing to similar measurements on $MnSb_2Te_4$ and $MnBi_2Te_4$ (Figs. S5 and S6 of the SM), we see that electrical transport in Sb-rich compositions are much more influenced by the magnetic order, likely because the Mn orbitals are located in the valence band and couple more closely to the hole carriers in $MnSb_{1.8}Bi_{0.2}Te_4$ and $MnSb_2Te_4$ [31,33].

Finally, we report on the appearance of an excess anomalous Hall signal in $MnSb_{1.8}Bi_{0.2}Te_4$ that is beyond the conventional AHE. This signal concentrates in the circled areas in the $\Delta R_{xy}(H)$ plot shown in Fig. 1(c). Following the literature [11,13,14], we fit the conventional AHE component $R_{xy}^A$ with a Langevin function (blue dashed line in Fig. 4(a)) and use $R_{xy}^T$ to denote the excess signal (green shaded area). Figures 4(b) and (c) plot $R_{xy}^T(H)$ obtained at several temperatures and different tilt angles of the external field respectively. The raw $\Delta R_{xy}(H)$ plots are given in Fig. S7 of the SM. The magnitude of $R_{xy}^T(H)$ decreases rapidly with increasing temperature and vanishes at $T > 10$ K. The signal peaks at $H \sim \pm 0.7$ T and persists to several Tesla. Figure 4(d) plots the angle dependence of the peak value $R_{peak}^T$. $R_{peak}^T(\theta)$ is non-monotonic and reaches a maximum of 0.1 Ω around $\theta \sim 60°$. The value corresponds to a fictitious field of $H_{eff} = R_{peak}^T \cdot ne \sim 1$ Tesla, which is similar to the strength of the fictitious field generated by skyrmionics spin textures in FeGe [9], i.e. the excess anomalous Hall signal in our $MnSb_{1.8}Bi_{0.2}Te_4$ sample has considerable strength.

Its appearance raises intriguing possibilities. The excess signal resembles the topological Hall effect (THE) discussed in the literature, which is commonly associated with excess Berry curvature caused by spin chirality generated by the Dzyaloshinskii-Moria interaction (DMI) [4-9]. In the bulk, the presence of a DMI term requires a noncentrosymmetric structure, a condition not met in an ideal $Mn(Sb,Bi)_2Te_4$ crystal. DMI can also occur at heterointerfaces combining magnetism and strong spin orbit coupling [10-12]. An inhomogeneous magnetization in our sample could inadvertently produce internal FM/TI interfaces that may host DMI interactions. Frustrated magnetic phases can offer alternative routes of generating non-collinear magnetic texture and unusual AHE [41]. Studies aimed to identify the low-temperature magnetic order and structural symmetry of our $MnSb_{1.8}Bi_{0.2}Te_4$ sample can further elucidate the origin of the excess AHE signal.

In summary, we combine electrical transport, bulk magnetometry and neutron diffraction studies to show evidence of a ferromagnetic ground state with a Curie temperature of 26 K in $MnSb_{1.8}Bi_{0.2}Te_4$. Our work is an encouraging step towards realizing a ferromagnetic topological insulator. Its vdW geometry opens up possibilities of forming heterostructures and gating tuning. Studies that illuminate the synthesis conditions of different magnetic phases in the $Mn(Sb, Bi)_2Te_4$ family will greatly facilitate future explorations of their topological and magnetic properties.

**Figure Captions:**

Figure 1. (a) Schematic stacking order of Mn(Sb,Bi)$_2$Te$_4$. (b) An optical image of a typical MnSb$_{1.8}$Bi$_{0.2}$Te$_4$ device in a Hall bar geometry. (c) The Hall resistance $R_{xy}(H)$ on a MnSb$_{1.8}$Bi$_{0.2}$Te$_4$ device at selected temperatures as labeled in the plot. Arrows indicate the field sweep direction. The green dashed line illustrates the zero-field slope $dR_{xy}/dH$. The inset shows the full-range down sweep of $R_{xy}(H)$ at 2 K. The black dashed line illustrates the high-field $dR_{xy}/dH$. ((d) $M(H)$ of a MnSb$_{1.8}$Bi$_{0.2}$Te$_4$ crystal. The coercive field $H_c$ = 310 Oe. The remnant magnetization $M_0$ = 0.6 $\mu_B$/Mn. Inset: $M(H)$ to 7 T showing a saturated magnetization of ~ 1.8 $\mu_B$/Mn. (e) The anomalous Hall component $\Delta R_{xy}(H)$ for the device shown in (c) at selected temperatures. $\Delta R_{xy}(H)$ is obtained by subtracting the normal Hall contribution $R_0 H$ from the measured $R_{xy}(H)$. The green dashed lines illustrate the process of obtaining the saturation field $H_s$ for the $T$ = 30 K trace. Up and down sweeps produce the same $H_s$.

Figure 2. (a) The zero-field slope $dR_{xy}/dH$ as a function of temperature in MnSb$_{1.8}$Bi$_{0.2}$Te$_4$ (magenta circles) and MnSb$_2$Te$_4$ (blue squares). The dashed lines are guide to the eye. ((b) The temperature-dependent magnetic susceptibility $M/H$ of a MnSb$_{1.8}$Bi$_{0.2}$Te$_4$ sample measured at three external fields $H$ = 50, 100, and 1000 Oe under both ZFC and FC conditions as labeled in the graph. $H$ // $c$. (c) and (d) The main panels show the temperature-dependent elastic neutron scattering centered at the (1 0 1) reflection in MnSb$_{1.8}$Bi$_{0.2}$Te$_4$ (c) and the (1 0 2.5) reflection in MnSb$_2$Te$_4$ (d). Mean-field fits (solid lines) yield $T_c$ = 26.3 K in (c) and $T_N$ = 19.5 K in (d). The insets show scans along the (1 0 L) direction at 5 K and 45 K with the * symbol marking the (1 0 1) reflection in (c) and the (1 0 2.5) reflection in (d). Error bars in (c) and (d) are given by standard deviations of the Poisson distribution.

Figure 3. (a) Moment saturation field $H_s$ vs $T$ in MnSb$_{1.8}$Bi$_{0.2}$Te$_4$ obtained from $\Delta R_{xy}(H)$ data shown in Fig. 1(e). Solid symbols are data below $T_C$. Hollow symbols are data above $T_C$ and use the right axis. (b) Lower panel: Temperature-dependent magnetoresistance $R_{xx}(T)$ taken at fixed magnetic field as labeled in the plot. The black dashed lines divide the curves into three regions according to the sign of $dR/dT$. The boundary points are plotted in the upper panel of (b). The symbols follow the notation of (a). (c) Normalized magnetoresistance MR = $[R_{xx}(H) - R_{xx}(0)]/R_{xx}(0) \times 100\%$ at selected temperatures. Arrows indicate field sweep direction.

Figure 4. Topological Hall effect in MnSb$_{1.8}$Bi$_{0.2}$Te$_4$. (a) $R_{xy}^{A+T}(H)$, which is the same as $\Delta R_{xy}(H)$ in Fig. 1(e) at $T$ = 2 K and with a tilt angle $\theta$ = 56° as illustrated in the inset. The red dashed line is a Langevin fit to the anomalous Hall effect with an uncertainty of 0.9 m$\Omega$. The area shaded in green indicates the excess Hall effect contribution $R_{xy}^T$. (b) $R_{xy}^T(H)$ at $\theta$ = 0° and $T$ = 2, 5, and 10 K. Data obtained from up and down sweeps are shifted horizontally to coincide at $H$ = 0. (c) $R_{xy}^T(H)$ at $T$ = 2 K and selected tilt angle $\theta$. $R_{peak}^T$ is marked by the * symbol. (d) The angular dependence of $R_{peak}^T$ at $T$ = 2 K. The right axis labels the effective magnetic field $H_{eff}$ = $R_{peak}^T/R_0$.

**Acknowledgement**
Y. C., Y-W. C., K. H. and J. Z. are supported by NSF through NSF-DMR-1708972. Y. C. also acknowledges support by the China Scholarship Council. Support for crystal growth and characterization was provided by the National Science Foundation through the Penn State 2D Crystal Consortium-Materials Innovation Platform (2DCC-MIP) under NSF cooperative agreement DMR-1539916. Z.Q.M. also acknowledges the support of NSF-DMR-1707502. We thank Xia Hong and Cui-Zu Chang for helpful discussions. P. Z. would like to thank Prof. J. Kusz for help with Single Crystal measurements and the support of NIST through the Guest Researcher Program.

**References**
[1] M. Z. Hasan and C. L. Kane, Rev Mod Phys **82**, 3045 (2010).
[2] C.-Z. Chang *et al.*, Science **340**, 167 (2013).
[3] C.-X. Liu, S.-C. Zhang, and X.-L. Qi, Annual Review of Condensed Matter Physics **7**, 301 (2016).
[4] N. Nagaosa and Y. Tokura, Nature Nanotechnology **8**, 899 (2013).
[5] A. Fert, N. Reyren, and V. Cros, Nature Reviews Materials **2**, 17031 (2017).
[6] Y. Taguchi, Y. Oohara, H. Yoshizawa, N. Nagaosa, and Y. Tokura, Science **291**, 2573 (2001).
[7] A. Neubauer, C. Pfleiderer, B. Binz, A. Rosch, R. Ritz, P. G. Niklowitz, and P. Boni, Phys Rev Lett **102** (2009).
[8] X. Z. Yu, N. Kanazawa, Y. Onose, K. Kimoto, W. Z. Zhang, S. Ishiwata, Y. Matsui, and Y. Tokura, Nature Materials **10**, 106 (2011).
[9] S. X. Huang and C. L. Chien, Phys Rev Lett **108**, 267201 (2012).
[10] S. Heinze, K. von Bergmann, M. Menzel, J. Brede, A. Kubetzka, R. Wiesendanger, G. Bihlmayer, and S. Blügel, Nature Physics **7**, 713 (2011).
[11] J. Matsuno, N. Ogawa, K. Yasuda, F. Kagawa, W. Koshibae, N. Nagaosa, Y. Tokura, and M. Kawasaki, Science Advances **2**, e1600304 (2016).
[12] J. Jiang *et al.*, arXiv preprint arXiv:1901.07611 (2019).
[13] J. Gallagher, K. Meng, J. Brangham, H. Wang, B. Esser, D. McComb, and F. Yang, Phys Rev Lett **118**, 027201 (2017).
[14] W. Wang *et al.*, Nature Materials **18**, 1054 (2019).
[15] B. Huang *et al.*, Nature **546**, 270 (2017).
[16] C. Gong *et al.*, Nature **546**, 265 (2017).
[17] S. Jiang, L. Li, Z. Wang, K. F. Mak, and J. Shan, Nature Nanotechnology **13**, 549 (2018).
[18] T. Song *et al.*, Science **360**, 1214 (2018).
[19] Y. Deng *et al.*, Nature **563**, 94 (2018).
[20] W. Xing, L. Qiu, X. Wang, Y. Yao, Y. Ma, R. Cai, S. Jia, X. C. Xie, and W. Han, Physical Review X **9**, 011026 (2019).
[21] M. M. Otrokov *et al.*, arXiv preprint arXiv:1809.07389 (2018).
[22] M. M. Otrokov *et al.*, Phys Rev Lett **122**, 107202 (2019).
[23] J. Cui, M. Shi, H. Wang, F. Yu, T. Wu, X. Luo, J. Ying, and X. Chen, Physical Review B **99**, 155125 (2019).
[24] Y. Deng, Y. Yu, M. Z. Shi, J. Wang, X. H. Chen, and Y. Zhang, arXiv preprint arXiv:1904.11468 (2019).

[25] C. Liu *et al.*, arXiv preprint arXiv:1905.00715 (2019).
[26] D. Zhang, M. Shi, T. Zhu, D. Xing, H. Zhang, and J. Wang, Phys Rev Lett **122**, 206401 (2019).
[27] Y. Gong *et al.*, Chinese Physics Letters **36**, 076801 (2019).
[28] J. Li, Y. Li, S. Du, Z. Wang, B.-L. Gu, S.-C. Zhang, K. He, W. Duan, and Y. Xu, Science Advances **5**, eaaw5685 (2019).
[29] J. Ge, Y. Liu, J. Li, H. Li, T. Luo, Y. Wu, Y. Xu, and J. Wang, arXiv preprint arXiv:1907.09947 (2019).
[30] S. H. Lee *et al.*, Physical Review Research **1**, 012011 (2019).
[31] J. Q. Yan, S. Okamoto, M. A. McGuire, A. F. May, R. J. McQueeney, and B. C. Sales, Physical Review B **100**, 104409 (2019).
[32] J. Q. Yan *et al.*, Physical Review Materials **3**, 064202 (2019).
[33] S. V. Eremeev, M. M. Otrokov, and E. V. Chulkov, Journal of Alloys and Compounds **709**, 172 (2017).
[34] R. C. Vidal *et al.*, aXiv:1906.08394 (2019).
[35] C. Hu *et al.*, arXiv preprint arXiv:1905.02154 (2019).
[36] N. Nagaosa, J. Sinova, S. Onoda, A. H. MacDonald, and N. P. Ong, Rev Mod Phys **82**, 1539 (2010).
[37] Y. Moritomo, Y. Tomioka, A. Asamitsu, Y. Tokura, and Y. Matsui, Physical Review B **51**, 3297 (1995).
[38] J. S. Gardner, M. J. P. Gingras, and J. E. Greedan, Rev Mod Phys **82**, 53 (2010).
[39] J. Wu and C. Leighton, Physical Review B **67**, 174408 (2003).
[40] J. D. Bocarsly, C. Heikes, C. M. Brown, S. D. Wilson, and R. Seshadri, Physical Review Materials **3**, 014402 (2019).
[41] D. Boldrin and A. S. Wills, Advances in Condensed Matter Physics **2012**, 1 (2012).

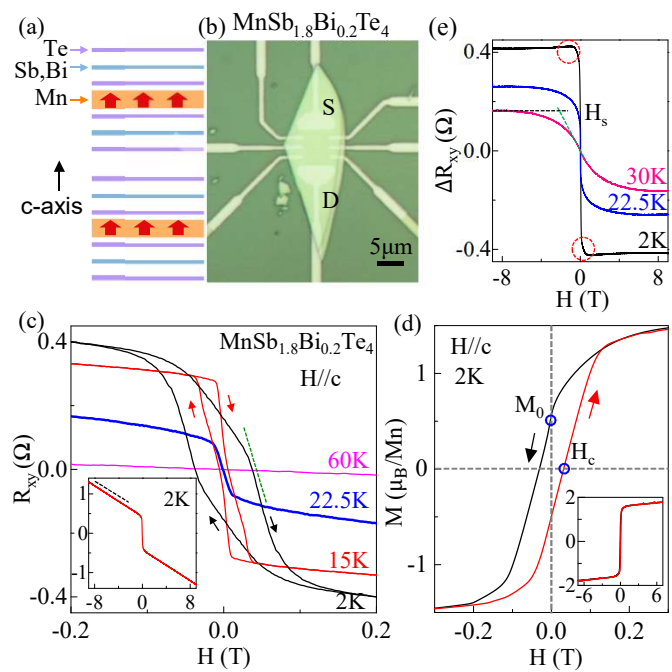

Figure 1

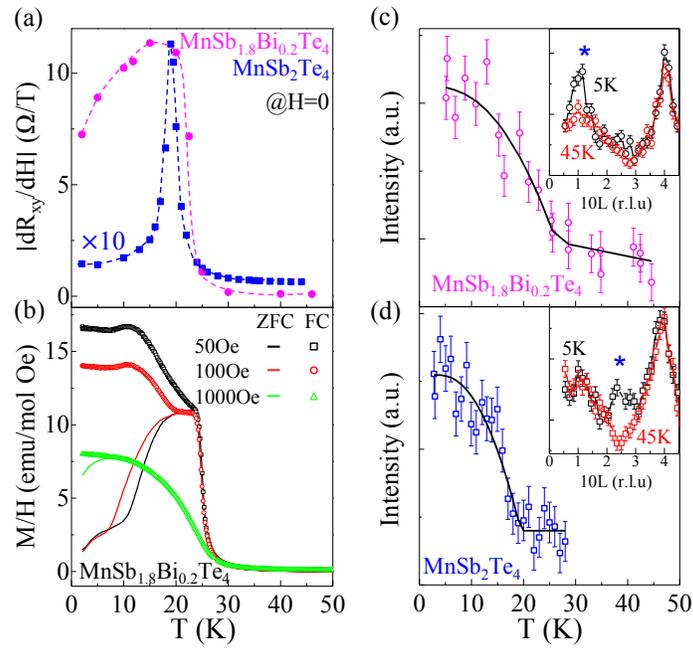

Figure 2

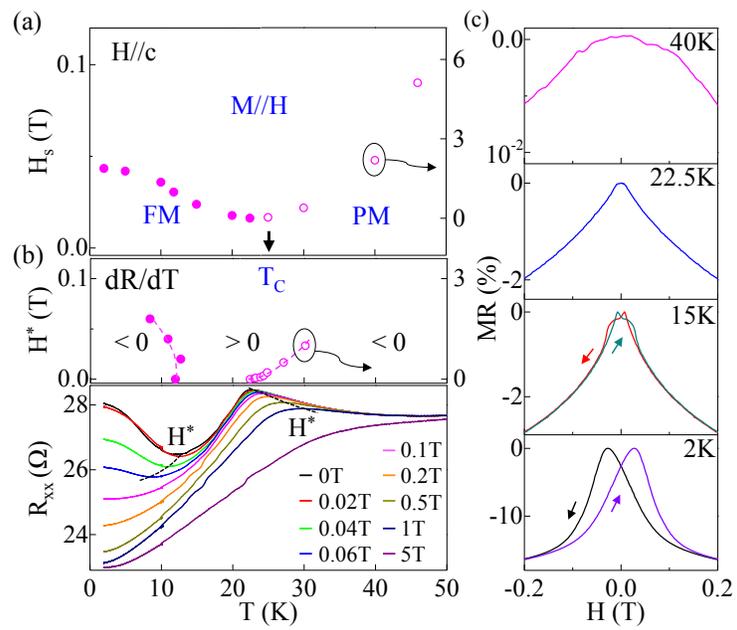

Figure 3

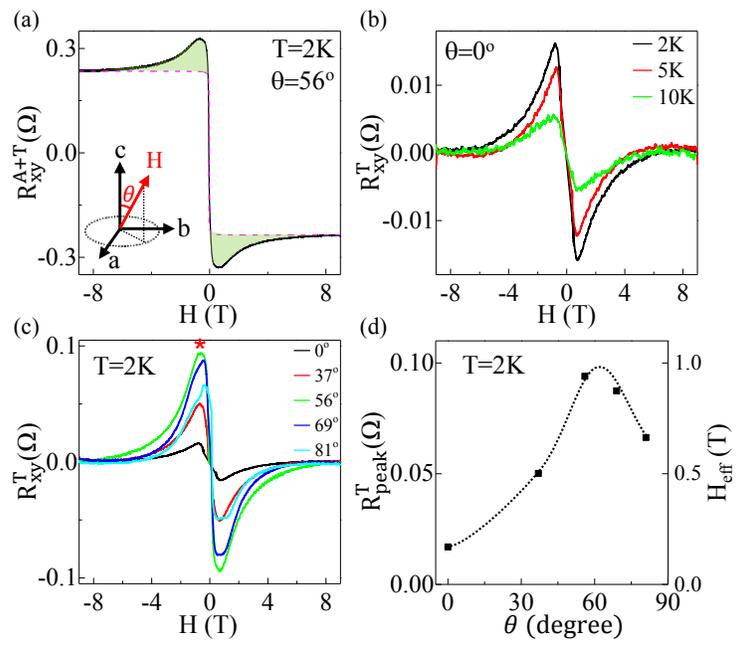

Figure 4

# Supplementary Material for

# Ferromagnetic van der Waals compound MnSb$_{1.8}$Bi$_{0.2}$Te$_4$

Yangyang Chen[1,3], Ya-Wen Chuang[1], Seng Huat Lee[1,2], Yanglin Zhu[1], Kevin Honz[1], Yingdong Guan[1], Yu Wang[1], Ke Wang[5], Zhiqiang Mao[1,2] and Jun Zhu[1]

Colin Heikes[4], P. Quarterman[4]
Pawel Zajdel[6], Julie A. Borchers[4]
William Ratcliff II[4,7]

[1]*Department of Physics, The Pennsylvania State University, University Park, Pennsylvania 16802, USA*
[2]*2DCC, The Pennsylvania State University, University Park, Pennsylvania 16802, USA*
[3] *International Center for Quantum Materials, School of Physics, Peking University, Beijing 100871, PR China*
[4]*NIST Center for Neutron Research, NIST, Gaithersburg, MD 20899, USA*
[5]*Materials Characterization Laboratory, Materials Research Institute, The Pennsylvania State University, University Park, Pennsylvania 16802, USA*
[6] *Institute of Physics, University of Silesia, ul. 75 Pulku Piechoty 1, 41-500, Chorzow, Poland*
[7]*Department of Materials Science and Engineering, University of Maryland, College Park, MD 20742, USA*

**Online Supplementary Material Content**

1. Synthesis and characterizations of MnSb$_{1.8}$Bi$_{0.2}$Te$_4$ and MnSb$_2$Te$_4$ bulk single crystals
2. The normal Hall effect and carrier density in MnSb$_{1.8}$Bi$_{0.2}$Te$_4$ and MnSb$_2$Te$_4$
3. Characterization of the hysteresis and remnant R$_{xy}$ in MnSb$_{1.8}$Bi$_{0.2}$Te$_4$
4. Temperature-dependent transport and magneto-transport in MnSb$_2$Te$_4$
5. Temperature-dependent transport and magneto-transport in MnBi$_2$Te$_4$
6. Anomalous Hall effect in MnSb$_{1.8}$Bi$_{0.2}$Te$_4$

**Section 1: Synthesis and characterizations of MnSb$_{1.8}$Bi$_{0.2}$Te$_4$ and MnSb$_2$Te$_4$ bulk single crystals**

MnSb$_{1.8}$Bi$_{0.2}$Te$_4$ and MnSb$_2$Te$_4$ single crystals were synthesized using a self-flux method similar to that used by Yan J.-Q. *et al.*, (arXiv:1905.00400). High purity of Mn powder (99.95%), Bi shot (99.999%), antimony shot (99.9999%) and Te ingot (99.9999+%) were mixed with the molar ratio of Mn:Sb:Bi:Te = 1:9:1:16 for MnSb$_{1.8}$Bi$_{0.2}$Te$_4$ and 1:10:0:16 for MnSb$_2$Te$_4$ and were loaded in an Al$_2$O$_3$ crucible and then sealed in evacuated quartz tubes. The mixtures were heated up to 900°C for 12 hours to promote homogeneous melting and followed by a slow cooling down to around 630 °C with a rate of 2 °C/hour. The ampoules were then quenched immediately, and the excessive (Bi,Sb)$_2$Te$_3$ flux was removed by centrifugation. We chose the quenching temperature of 630 °C considering that the melting point of Sb$_2$Te$_3$ is 620 °C. Two different Muffle furnaces were used to synthesize the two different compositions. Although the quenching temperatures for all our growth are nominally the same, the actual quenching temperatures each ampoule experiences could be slightly different because of the position of the ampoule and the spatially inhomogeneous temperature profile inside a Muffle furnace. In addition, the cooling rate of each ampoule during the quenching process may also be slightly different, since it depends on how fast the ampoule is removed from the furnace and loaded into the centrifuge. As a result, even crystals grown in the same Muffle furnace can experience different quenching temperature and cooling rate, which may lead to the stabilization of different metastable phases. We have observed distinctly different types of disorder in our MnSb$_{1.8}$Bi$_{0.2}$Te$_4$ and MnSb$_2$Te$_4$ crystals, as manifested in microscopy studies discussed below.

The good crystallinity of the MnSb$_{1.8}$Bi$_{0.2}$Te$_4$ and MnSb$_2$Te$_4$ single crystals was confirmed by the sharp (*0 0 L*) x-ray diffraction (XRD) peaks, as shown in Fig. S1. Both (*0 0 L*) series diffraction patterns look similar to those seen in MnBi$_2$Te$_4$, suggesting the Sb-rich samples also crystallize in a rhombohedral structure with the space group $R\bar{3}m$. A minor secondary phase (< 5%) of (Bi/Sb)$_2$Te$_3$ is also observed on some flakes, suggesting the intergrowth of (Bi/Sb)$_2$Te$_3$. Atomic-resolution high-angle annular dark-field (HAADF)-STEM image taken on both MnSb$_{1.8}$Bi$_{0.2}$Te$_4$ and MnSb$_2$Te$_4$ samples show high crystallinity and clear rhombohedral stacking of the Te-Sb/Bi-Te-Mn-Te-Sb/Bi-Te septuple layers (SLs) along the *c* axis. Stacking faults are observed in MnSb$_2$Te$_4$ (Fig. S2 (b)) but not in MnSb$_{1.8}$Bi$_{0.2}$Te$_4$ (Fig. S2(a)). Considerable anti-site defects are present in both compositions. Their potential contribution to the magnetic ground state of the compounds requires more in-depth structural analysis to clarify.

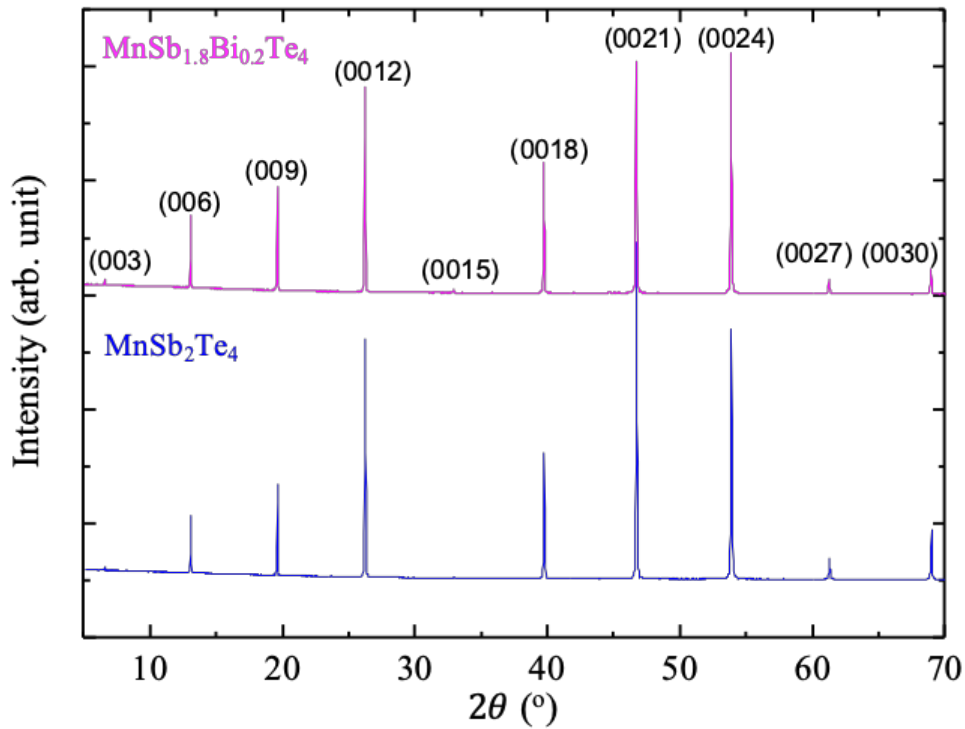

Fig. S1. X-ray diffraction pattern (XRD) of MnSb$_{1.8}$Bi$_{0.2}$Te$_4$ (magenta) and MnSb$_2$Te$_4$ (blue) bulk single crystals showing a series of sharp (00L) peaks. They match well with the ICDD data base PDF card 04-020-8214 for MnSb$_{1.8}$Bi$_{0.2}$Te$_4$ and card 04-020-8214 for MnSb$_2$Te$_4$, demonstrating excellent crystallinity.

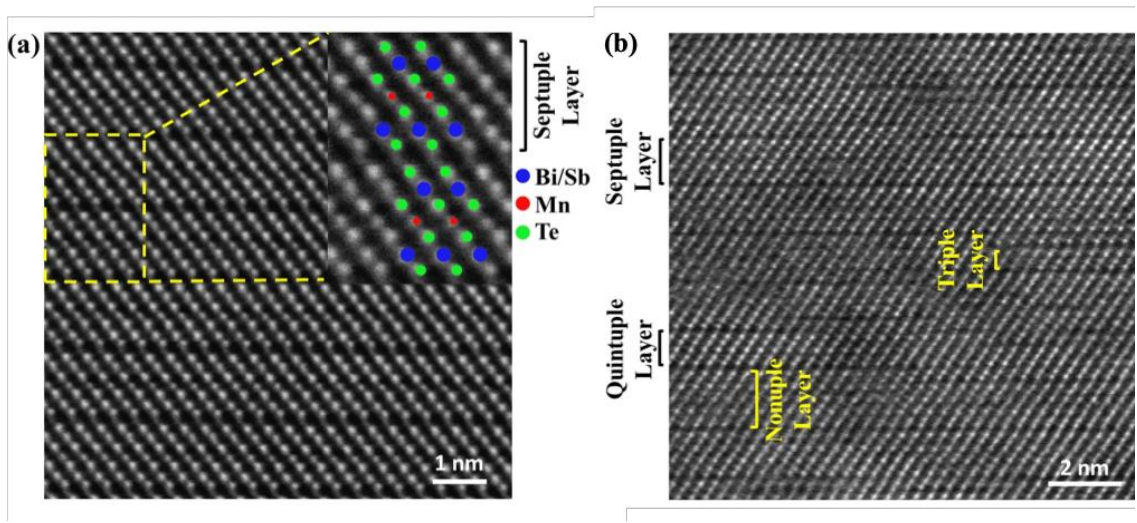

Fig. S2. HAADF-STEM image of of MnSb$_{1.8}$Bi$_{0.2}$Te$_4$ (a) and MnSb$_2$Te$_4$ (b) taken from the [100] zone axis. Inset of (a): Magnified image of a one-unit cell with the atoms overlaid on top. Stacking faults are common in (b) but not in (a).

**Section 2: The normal Hall effect and carrier density in MnSb$_{1.8}$Bi$_{0.2}$Te$_4$ and MnSb$_2$Te$_4$**

As the inset to Fig. 1(c) shows, we determine the carrier density in our multi-layer devices using the slope of the Hall resistance $R_0 = dR_{xy}/dH$ near $H = 8$ T. The results are presented in Fig. S3(a) for a MnSb$_{1.8}$Bi$_{0.2}$Te$_4$ and a MnSb$_2$Te$_4$ device as a function of temperature. The carrier doping is p-type in both materials and is roughly independent of temperature in the range we probed. From these measurements, we estimate a hole density of a few $\times$ 10$^{13}$/cm$^2$/septuple layer. This means the Fermi level is in the bulk valence band of the materials. Though many layers are present in our devices, the linear dependence of $R_0$ on cos $\theta$ given in Fig. S3(b) shows that the electronic transport of the carriers are two dimensional in nature.

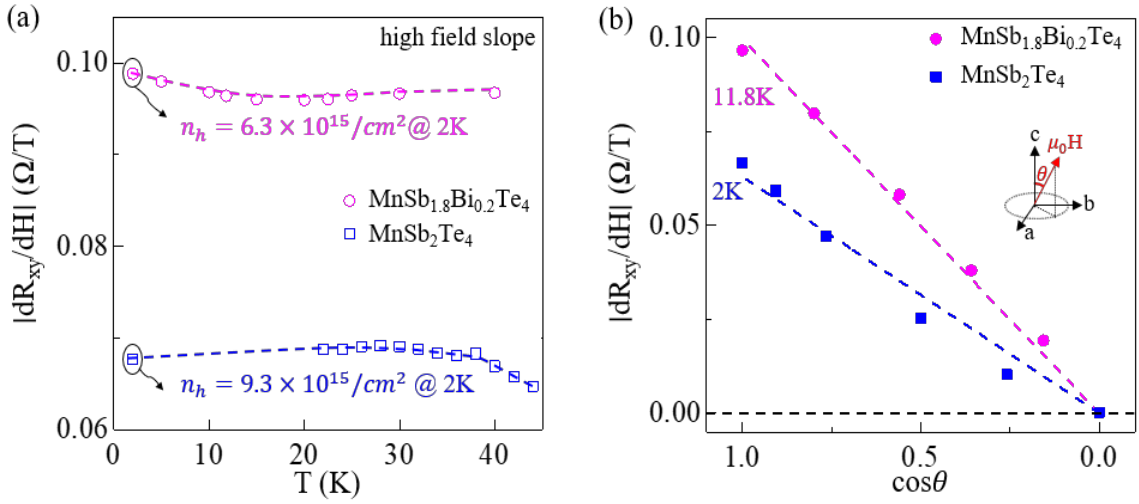

Fig. S3. (a) High-field slope $R_0 = dR_{xy}/dH$ in MnSb$_{1.8}$Bi$_{0.2}$Te$_4$ and MnSb$_2$Te$_4$ as a function of temperature. $R_0 = 1/ne$ yields the 2D carrier density. (b) The angle $\theta$ dependence of $R_0$. The inset illustrates the definition of $\theta$. The linear relation between $dR_{xy}/dH$ and cos $\theta$ indicates the 2D nature of the mobile carriers.

**Section 3. Characterization of the hysteresis and remnant $R_{xy}$ in MnSb$_{1.8}$Bi$_{0.2}$Te$_4$**

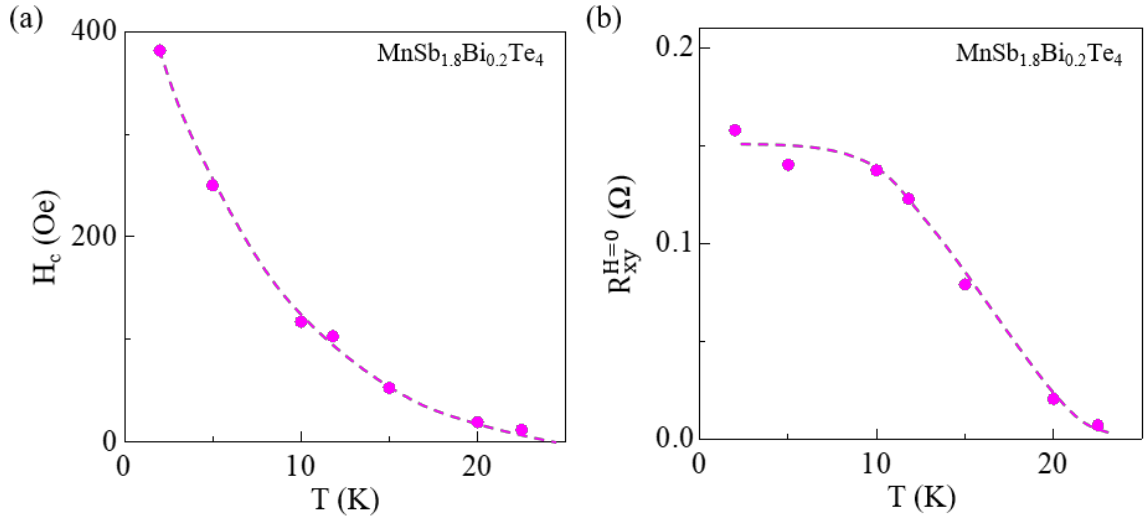

Fig. S4. The coercive field H$_c$ (a) and the remnant Hall resistance $R_{xy}^{H=0}$ (b) as a function of temperature extracted from the R$_{xy}$ data shown in Fig. 1(c). Both support the onset of a ferromagnetic order at T$_c$ > 20 K. Dashed lines are guide to the eye.

**Section 4. Temperature-dependent transport and magneto-transport in MnSb$_2$Te$_4$**

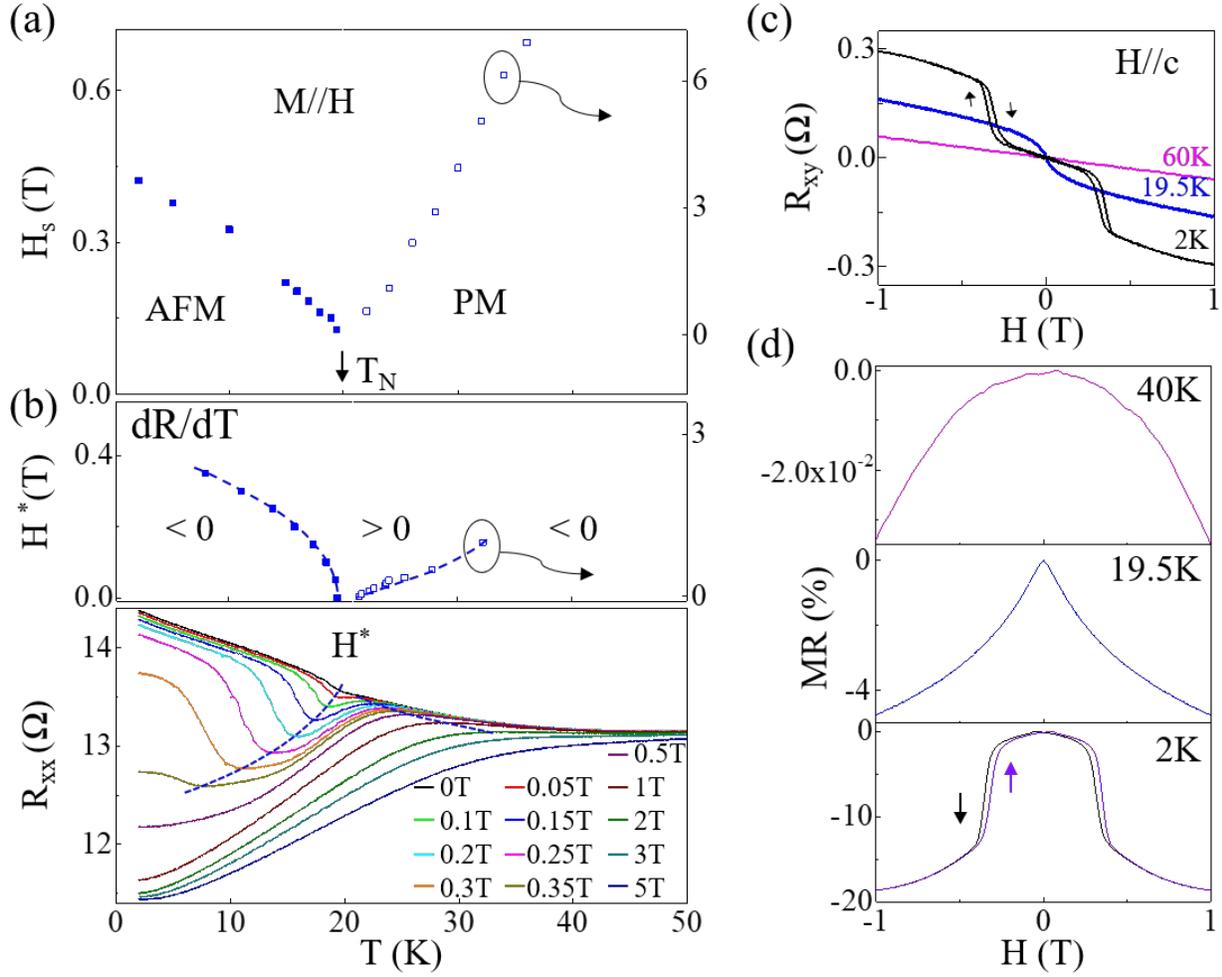

Fig. S5. Temperature-dependent resistance, $R_{xy}$ and magneto-resistance data on a MnSb$_2$Te$_4$ device. The notations of the figures follow Figs. 1 and 3 of the text on MnSb$_{1.8}$Bi$_{0.2}$Te$_4$. As the $H_s$ - $T$ diagram in (a) shows, the saturation field is 0.42 T in MnSb$_2$Te$_4$, compared to 0.04 T in MnSb$_{1.8}$Bi$_{0.2}$Te$_4$ (Fig. 3) and more than 7 T in MnBi$_2$Te$_4$ (Fig. S6). The two orders of magnitude variation in $H_s$ point to very different magnetic order in these three materials.

**Section 5. Temperature-dependent transport and magneto-transport in MnBi$_2$Te$_4$**

In this section, we provide in comparison data obtained on our MnBi$_2$Te$_4$ devices exfoliated from crystals grown in Ref. [1]. Our data are in very good agreement with results obtained by other groups on this material [1-3].

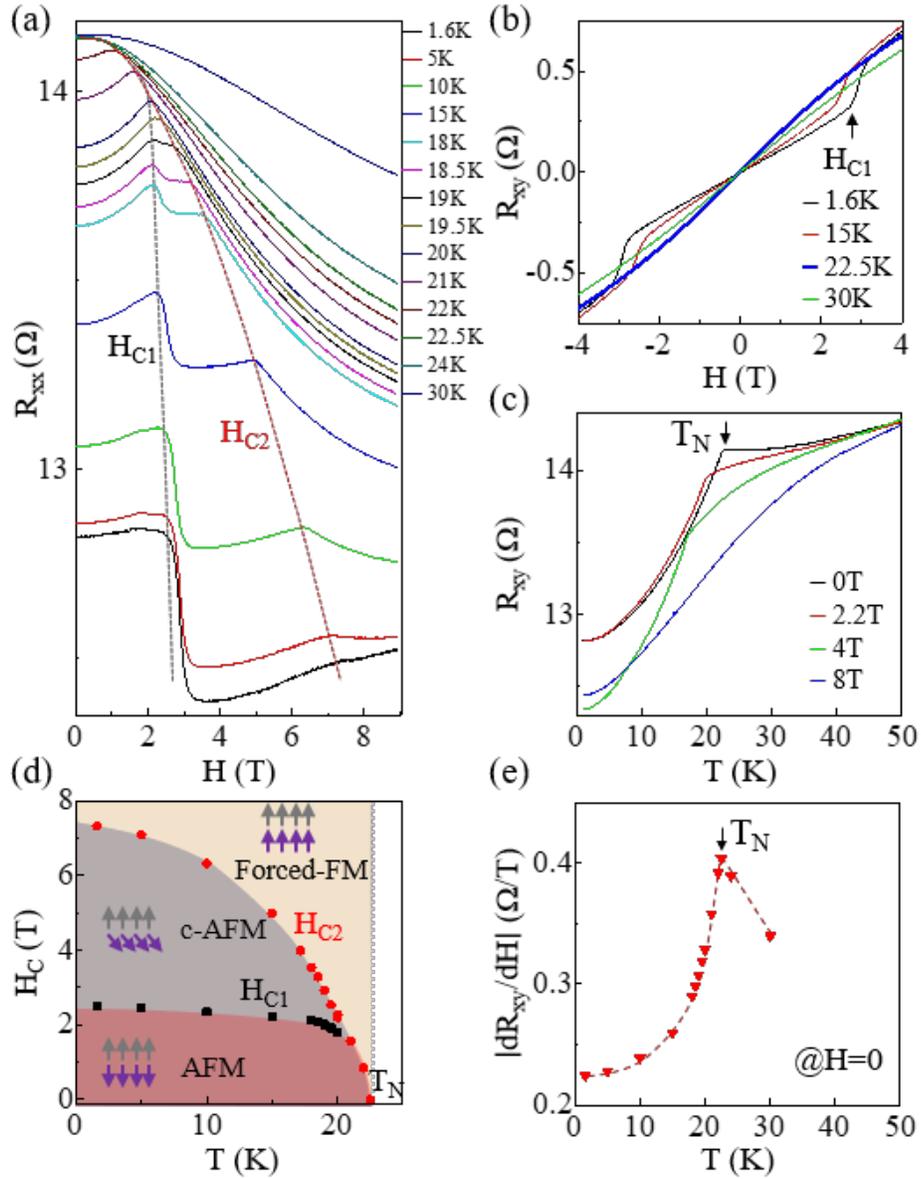

Fig. S6. Magnetoresistance $R_{xx}(H)$ (a) and Hall resistance $R_{xy}(H)$ at selected temperatures as labeled in the plots. A spin-flop transition is seen at $H_{C1}$, where both $R_{xx}$ and $R_{xy}$ undergo an abrupt jump. A second kink is seen at $H_{C2}$, which initiates a second rapid drop of $R_{xx}$. (c) The $T$-dependent $R_{xx}$ at selected fields. A cusp is visible in the black, red and green traces when the system is in an AFM or canted AFM phase. It disappears after the moments are aligned with the external field. (d) plots the phase diagram extracted

from the data in (a). (e) plots the zero-field slope $dR_{xy}/dH$ obtained from the Hall resistance data. It strongly resembles magnetic susceptibility measured on bulk crystals by us and others [1-3]. A cusp is seen at the Neel temperature $T_N$ = 22.5 K. $H // c$ in all measurements.

## Section 6. Anomalous Hall effect in MnSb$_{1.8}$Bi$_{0.2}$Te$_4$

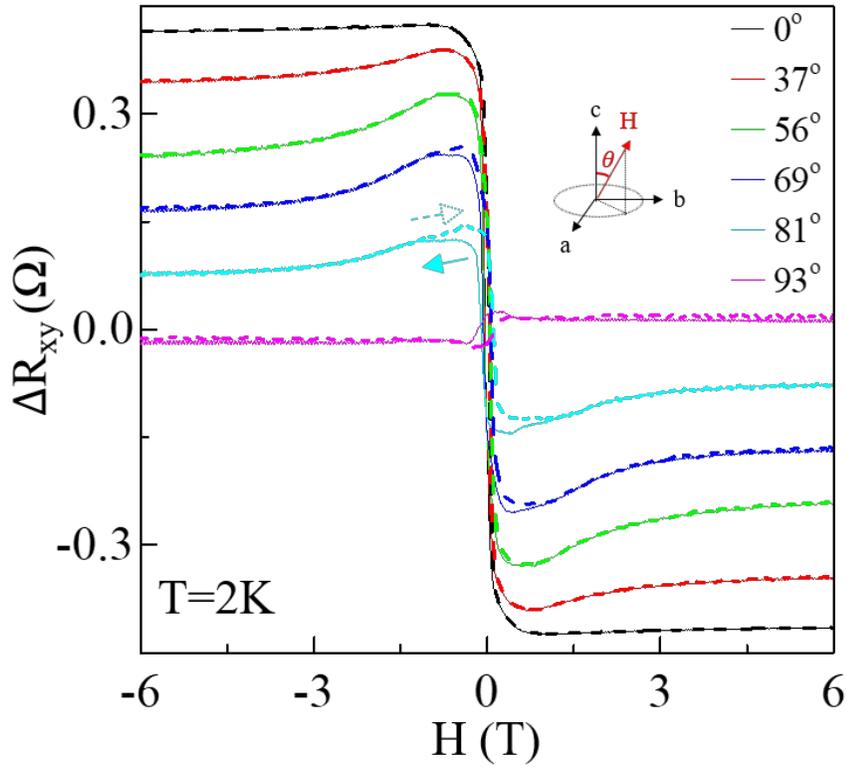

Fig. S7. $\Delta R_{xy}(H)$ of the MnSb$_{1.8}$Bi$_{0.2}$Te$_4$ device shown in (c) at different tilt angles as labeled in the plot. Solid and dashed lines correspond to down and upsweeps respectively. $T$ = 2 K.

**References**

[1]   S. H. Lee *et al.*, Physical Review Research **1**, 012011 (2019).
[2]   M. M. Otrokov *et al.*, arXiv preprint arXiv:1809.07389  (2018).
[3]   J. Cui, M. Shi, H. Wang, F. Yu, T. Wu, X. Luo, J. Ying, and X. Chen, Physical Review B **99**, 155125 (2019).